# Increasing Quantum Communication Rates Using Hyperentangled Photonic States

**Liat Nemirovsky Levy[1], Uzi Pereg[2], and Mordechai Segev[1,2]***

[1] *Physics Department and Solid-State Institute, Technion, Haifa 3200003, Israel*
[2] *Department of Electrical Engineering, Technion, Haifa 3200003, Israel*
*[msegev@technion.ac.il](mailto:msegev@technion.ac.il)

**Abstract**

Quantum communication is based on the generation of quantum states and exploitation of quantum resources for communication protocols. Currently, photons are considered as the optimal carrier of information, because they enable long-distance transition with resilience to decoherence, and they are relatively easy to create and detect. Entanglement is a fundamental resource for quantum communication and information processing, and it is of particular importance for quantum repeaters [1]. Hyperentanglement [2], a state where parties are entangled with two or more degrees of freedom (DoFs), provides an important additional resource because it increases data rates and enhances error resilience. However, in photonics, the channel capacity, i.e. the ultimate throughput, is fundamentally limited when dealing with linear elements. We propose a technique for achieving higher transmission rates for quantum communication by using hyperentangled states, based on multiplexing multiple DoFs on a single photon, transmitting the photon, and eventually demultiplexing the DoFs to different photons at the destination, using a Bell state measurement. Following our scheme, one can generate two entangled qubit pairs by sending only a single photon. The proposed transmission scheme lays the groundwork for novel quantum communication protocols with higher transmission rate and refined control over scalable quantum technologies.

Entanglement and quantum correlations between multipartite quantum sub-systems have recently attracted significant attention in both fundamental and applied research areas based on Quantum Mechanics. Multipartite entanglement is the cornerstone of a range of applications, including quantum communications [3], quantum computation (including also the photonic one-way quantum computing [4–6]), and quantum cryptography [7,8]. Increasing the dimensionality of entanglement by entangling more than two DoF ("hyperentanglement") is a key enabler for encoding information in large quantities. This results in high-capacity quantum communication channels, where photons are the ultimate carriers of information, facilitating means to manipulate the entanglement. Since long-range communication in current systems is based on single-mode optical fibers, the prospect of encoding the quantum information on single photons (rather than on pairs of photons or clusters) makes them attractive, because this will increase the channel capacity. The volume of information carried by single photons can be potentially enormous.

Photon-based quantum information uses the DoFs of light, where qubits are implemented on any physical states that can be regarded as two-level states. For example, spin angular momentum (SAM), spatial distribution of the light beam (e.g., orbital angular momentum (OAM) encoding) [9,10], path encoding (propagation direction or k-vector), time [11] (time-bin and time-frequency encoding) [12–14] etc. In standard settings, entanglement is implemented on pairs of photons, where the DoF is either polarization or path. However, in principle, any binary quantum alternative can represent a qubit [15]. That is, a single quantum particle may naturally possess multiple degrees of freedom (DoFs) and thus several qubits can be encoded on a single particle. Such a state is called a hyperentangled state [2], where hyperentanglement refers to entanglement with at least two DoFs. Hyperentanglement can be in principle implemented on a single particle [16–18] or between two (or more) parties entangled with several DoFs [9,19]. Hyperentanglement offers significant advantages in quantum communication protocols, e.g., secure superdense coding (SDC) [20] and cryptography [21]. For example, if the DoFs of a hyperentangled pair of particles can be considered as a qubit, then the state of the hyperentangled pair is a tensor product of Bell states in each of the $n \otimes n$ variables (where each represents an effective qubit encoded in the DoF). Thus, there are $4^n$ entangled states overall. This novel construction upgrades the quantum communication channel, as it doubles the information capacity since more qubits are multiplexed in each pair of particles. In this vein, expanding the hyperentanglement to even more DoFs offer higher information capacity. Namely, a large Hilbert space can be achieved through entanglement in more than one DoF, and the DoF can also be expanded to more than two dimensions (known as high-dimensional entanglement), such as OAM DoF [9], or frequency [22].

In recent years, multipartite entanglement for increasing the channel capacity has been proposed and demonstrated [23–27]. However, at present, the main obstacle in establishing large-scale quantum networks is inherent loss through the transmission channel [28]. Due to the ever-increasing demand for high-capacity channels, hyperentangled photonic states hold the greatest promise for an efficient implementation of various quantum communication protocols, providing a high capacity link for sending either classical or quantum information, by a more efficient encoding of each physical photon. Thus, even though there are still losses of photons, each photon contains a larger amount of information, which improves reliability and resilience against the loss. However, storing information on single photons also involves some limitations, since this means that the non-classical correlations manifesting the entanglement are always local. Thus, seemingly, by entangling multiple qubits on a single photon precludes the use of one of the main advantages of entanglement: the ability to measure the correlations between the qubits in separate locations.

Here, we show how to overcome the limitations of encoding multiple qubits on a single photon. We present a scheme for transmitting qubits at a higher rate by multiplexing $N$ qubits on a single photon via a quantum teleportation protocol [29], transmitting a string of single photons, and eventually - at the destination - demultiplexing the quantum information to $N$ photons each carrying a single qubit. This enables the information to be processed in parallel, exploiting the nonlocality of quantum information processing. The hyperentanglement enables multiplexing the quantum information as a composite state carried on a single photon, transmitting the single photon, and eventually demultiplexing the composite state from a single photon to multiple photons which enables parallel quantum processing.

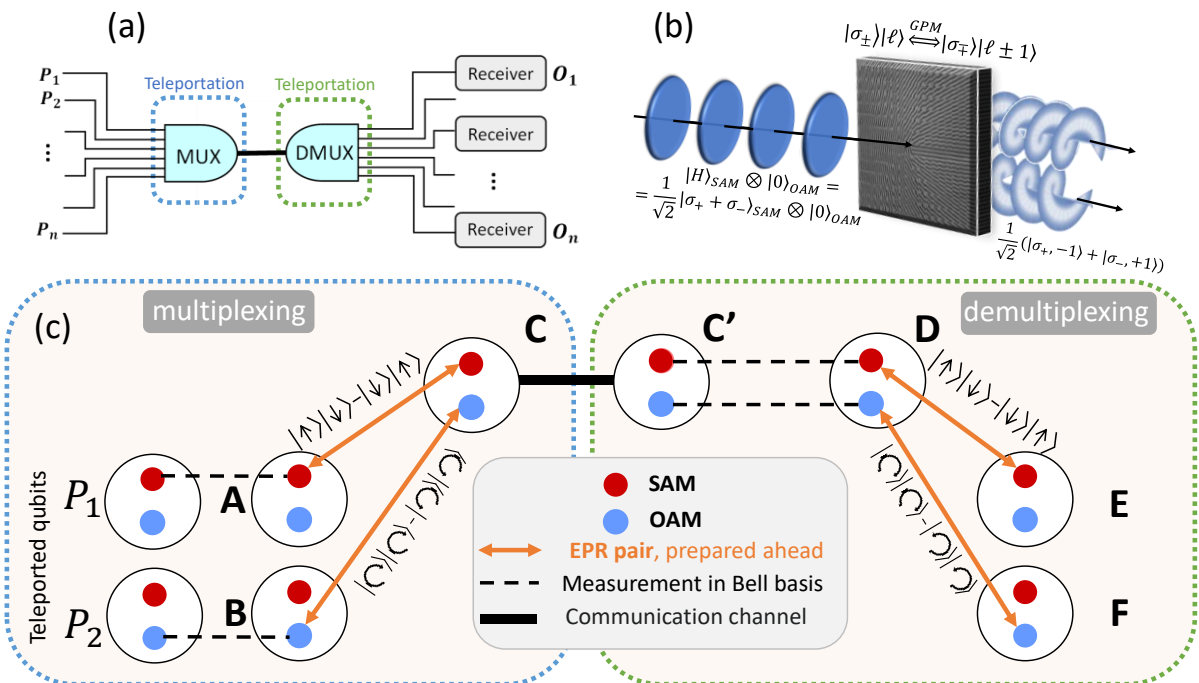

**Figure 1** **(a)** General scheme for the increasing communication rate with $N$ DoFs. MUX – multiplexing, the blue dashed line highlights the multiplexing process (using quantum teleportation) detailed in (c). Likewise, DMUX – demultiplexing, the green dashed line encompasses the process described in (c). **(b)** Example: encoding a qubit on a single photon. The metasurface that entangles between SAM and OAM of photons. The photon carries zero OAM before passing through the metasurface, after which it exits as a superposition of two circular polarizations, with the corresponding vortex phase fronts opposite to one another [22]. This metasurface is used to prepare the initial state for our protocol. **(C)** Scheme for multiplexing, transmitting and demultiplexing hyperentanglement using quantum teleportation. The multiplexing part (demultiplexing) is performed by teleporting SAM and OAM DoFs from separate photons (single photon) to a single photon (separate photons) and is labeled in dashed blue (green) line.

For our protocol, we use quantum teleportation [29], which provides a 'disembodied' way to transfer quantum states from one object to another at a distant location, assisted by previously shared entangled states and a classical communication channel. Recently, the teleportation of multiple DoFs in a single photon was demonstrated [30].

The technique presented here is general, not restricted to a specific implementation, applicable to any set of DoFs. As a concrete example, we propose to use DoFs of spin angular momentum (SAM) and orbital angular momentum (OAM). The technique has two parts (see Fig. 1(a)): a multiplexing part, where we teleport the SAM and the OAM of Photons $P_1$ and $P_2$, respectively, to Photon $C$, and a demultiplexing part- where we teleport again the DoFs from photon $C$ to separate photons $E$ and $F$. The encoding protocol comprises of Photon $C$ (the "information carrier") and Photons $A$ and $B$, which are entangled (separately) in their SAM and OAM with Photon $C$, respectively. The decoding protocol comprises of Photons $D$, $E$ and $F$.

The quantum teleportation protocol requires a previously shared entangled state. Since we perform two teleportation protocols (one for Photon $P_1$ and one for photon $P_2$), we need two pairs of entangled photons. To generate these two pairs, we use a metasurface (Fig. 1(b)). These are masks which imprint a different wavefunction to each polarization of the electromagnetic field. These devices enable local control of optical polarization, and were recently used to imprint entanglement between the SAM and the OAM of a photon [16].

To design a metasurface that entangles the photon SAM with OAM, the nanoantenna orientations are designed in such a way that the mask adds or subtracts $\Delta l = 1$ (one quanta of OAM), depending on the sign of the spin, and performs a spin-flip $|\sigma_+\rangle \leftrightarrow |\sigma_-\rangle$. Such a metasurface [16], presented in Fig. 1(b), performs the unitary transformation: $|\sigma_\pm\rangle|l\rangle \overset{GPM}{\Longleftrightarrow} |\sigma_\mp\rangle|l \pm \Delta l\rangle$, where $\sigma_\pm$ represents the spin states of the photon (right and left-handed circular polarizations), and $l$ represents the OAM of the photon.

In addition to the metasurface, our transmitter employs Photons $P_1$ and $P_2$. We are interested in locally teleporting the spin and the OAM of Photons $P_1$ and $P_2$, respectively, to Photon $C$ (as illustrated in Fig. 1(c)). To implement the teleportation protocol of both SAM and OAM to Photon $C$, the transmitter prepares ahead a hyperentangled photon pair with Photons $A$ and $B$, which are simultaneously entangled in both their SAM and their OAM with Photon $C$. The joint state of the transmitter's qubits can be expressed as

$$(|\sigma_+, +1\rangle_C \otimes |S_B, -1\rangle_B - |\sigma_+, -1\rangle_C \otimes |S_B, +1\rangle_B) \otimes |\sigma_-, L_A\rangle_A \\ - (|\sigma_-, +1\rangle_C \otimes |S_B, -1\rangle_B - |\sigma_-, -1\rangle_C \otimes |S_B, +1\rangle_B) \otimes |\sigma_+, L_A\rangle_A \quad (1)$$

where $\pm 1$ represent one quanta of OAM, $S_B$ is the SAM DoF of Photon B and $L_A$ is the OAM of Photon A. Equation (1) describes a state in which Photon $A$ and Photon $C$ are entangled in their spin, while Photons $B$ and $C$ are entangled in their OAM. Their combination yields a set of 16 hyperentangled Bell states.

Next, a necessary step in the local teleportation protocol of the transmitter is to perform a two-particle joint measurement of Photons $A$ and $P_1$, projecting them onto the 16-basis of orthogonal and complete hyperentangled Bell states, and discriminating one of them.

After the measurement in the Bell basis of Photons $A$ and $P_1$, Photon $C$ is projected onto the initial state of Photon $A$. The measurement results in the Bell basis of photons $A$ and $P_1$ and is encoded as four-bit classical information, which allows us to apply appropriate Pauli operations on Photon $C$ to perfectly reconstruct the initial spin of Photon $P_1$. This process is also performed with Photons $B$ and $P_2$, which are encoded with an OAM DoF. This scheme is presented by the blue dashed line in Fig. 1(c). This mechanism is an effective way to multiplex quantum information on a single photon. The single photon can now be transmitted to a desired destination (which could be far away, where the information can be demultiplexed into multiple photons, facilitating nonlocal quantum operations. The transmission is over a single communication channel –transmission of a single photon carrying two qubits, which doubles the quantum transmission rate.

To demultiplex the quantum information at the receiver, the local process of teleportation is repeated, separating the DoFs from Photon C' to Photons $E$ and $F$, demultiplexing the hyperentangled state. To teleport the spin and OAM DoFs in Photon $C'$, the receiver prepares ahead (as in the multiplexing teleportation protocol) a hyperentangled state consisting of 3 photons – D, E and F. This state is composed of two Bell states: Photons $D$ and $E$ which are entangled in their SAM, and another Bell pair of Photons $D$ and $F$ which are entangled in their OAM. Then, we measure the qubits encoded in Photons $C$ and $D$ in the Bell basis: One measurement is for the SAM of both photons, and the second is for the OAM of each photon. By measuring and encoding the result, the decoder can apply unitary gates on Photon $E$ and $F$, and perfectly reconstruct the teleported qubit. Thus, using only linear elements, the decoder has locally teleported a hyperentangled state from a single photon to two distinct photons. In a fashion similar to the multiplexing process, our decoder uses the teleportation of separate DoF to demultiplex the qubits to various photons (each DoF on a different photon). To be able to perform deterministic multiplexing and demultiplexing, one should be able to perform a complete Bell state measurement, which cannot be performed using "standard" EPR pairs and linear optics only. However, using hyperentanglement, one can actually isolate every individual Bell measurement (out of four possible outcomes) with a 100% certainty [31–33].

We show our reconstructed state in the simulation presented in Fig. 2. We simulate (using [34]) the protocol by sending two qubits of information using the transmission of a single photon through a noisy quantum channel. The most common noise in quantum optical communication protocols is the loss of photons. Therefore, to simulate the noise in the system, we assume an erasure channel – where a transmitter sends a qubit, and the receiver either receives the qubit correctly, or loses the qubit with some probability. With this in mind, we introduce photon loss noise into our system after each operation of the teleportation protocol presented in Fig. 1(c). Finally, to evaluate the performance of our system in the presence of noise, we measure the fidelity between the initial and the reconstructed states, while increasing the erasure rate (the probability that the photon is lost). The results are presented in Fig 2. The quantum capacity is the best ratio of information qubits per transmission, i.e., per photon. In the standard case of a single qubit per transmitted photon, the quantum capacity is given by $C_Q = 1 - 2\epsilon$, for $0 \leq \epsilon \leq \frac{1}{2}$, where $\epsilon$ is the erasure rate. Otherwise, if $\frac{1}{2} \leq \epsilon \leq 1$, i.e., more than half the

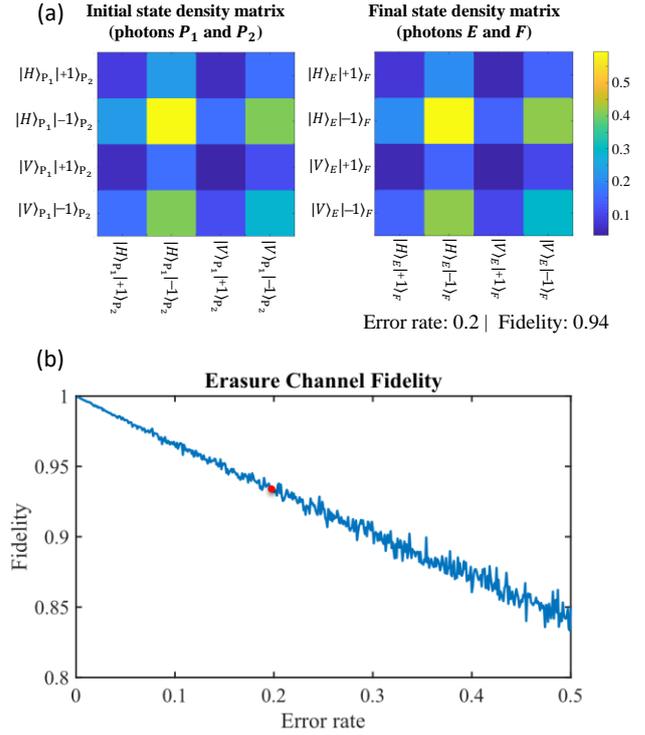

**Figure 2 (a)** Density matrices of the initial (before the multiplexing stage) and final (after the demultiplexing stage) states, with error rate (i.e., the chances of a photon to get lost) of 0.2 (red dot in Fig. 2(b)). The initial state is randomly generated (similar to the teleportation process, where the transmitted state is unknown). **(b)** Calculated fidelity as a function of error rate. Each point represents the average of 70 randomized states with the same error rate).

photons are lost to the environment, in which case the quantum channel capacity is zero. Using the standard approach, a single qubit is encoded on a single photon (namely, a single qubit per transmission). Our protocol improves the quantum capacity, i.e., the ratio of information qubits per transmission, by increasing the number of DoFs available for representing the qubits and multiplex them on a single photon (a single transmission). In our setting (sketched in Fig. 1(c)), the quantum capacity is doubled, hence $C_Q = 2(1 - 2\epsilon) \: for \: 0 \leq \epsilon \leq \frac{1}{2}$. We elaborate on the calculation of the quantum channel capacity in our setting in the supplementary information.

Thus far, we described the process in details for two DoFs. However, this protocol of multiplexing quantum information on a single photon, transmitting it, and eventually demultiplexing into multiple photons while recovering the quantum information in full, can be scaled up by increasing the size of Hilbert space of the transmitted qubits. We distinguish between two strategies to increase the Hilbert space: the first is adding $k$ DoFs as 2-level systems (i.e., adding more qubits); and the second is adding a DoF as an n-level system, where $n = 2^k$ (i.e., adding a qudit). From a purely mathematical perspective, the two options are equivalent. Both actions enlarge the Hilbert space, and enable multiplexing a greater amount of information on a single photon. Ideally, one can increase the Hilbert space by adding an indefinite number of DoFs and encoding all of them on a single photon, thereby increasing the quantum channel capacity by ($k$ or $n = 2^k$, respectively). However, there are practical limits on the number of different DoFs that can be carried by a single photon. For example, the number of detectors required for measuring is equal to the number of encoded qubits – not to the number of the physical photons. This sets a technological limit on the number of DoFs that can be encoded on a single photon.

Another method for increasing the dimensionality in the second strategy is by expanding the number of modes in the OAM DoF to more than two (i.e., encode a qudit in this DoF). For example, the metasurface can be designed to endow each polarization with multiple states of OAM, resulting in teleportation of high dimensional entangled photonic states. Another example, perhaps even more important, is employing the frequency as a DoF, where the number of possibilities can be enormous (the number of peaks in a frequency comb) [12]. Here, dispersive effects in the communication fiber may restrict the number of possibilities, but these are not conceptual and some of them might be accounted for in advance in the coding scheme.

As a case study, we can use two DoFs such as frequency and SAM. The SAM DoF is a two-level system, and the frequency DoF, as discussed earlier, can be extended to be N-dimensional. These two DoFs are orthogonal and can be encoded together on a single physical photon (i.e., the combined product channel). When encoding the DoFs together, the qubits transmission rate for general noise models can be larger than the sum of the quantum capacities for each one separately. This unique feature is known as the super-additivity of the quantum capacity. In addition, in some cases, we can even achieve super activation- where the quantum channel capacity of each channel (for each DoF) is zero, but the quantum channel capacity of the combined product channel takes a positive value [35]. In particular, this is the case when combining a symmetric erasure channel $\left(\epsilon = \frac{1}{2}\right)$, and an entanglement binding channel [36]. This feature is unique to quantum communication and does not occur in classical communication. A detailed discussion about the two schemes of increasing the Hilbert space of our qubits, (and by that increasing the rate of communication), is provided in the supplementary information.

One may also consider the task of sending classical information using the quantum channel, as in the simple setting of super-dense coding. The ultimate rate of classical information per quantum transmission is called the classical capacity of the quantum channel. Whether super-additivity applies to the classical capacity of a quantum product channel is yet an open problem [3] [37,38].

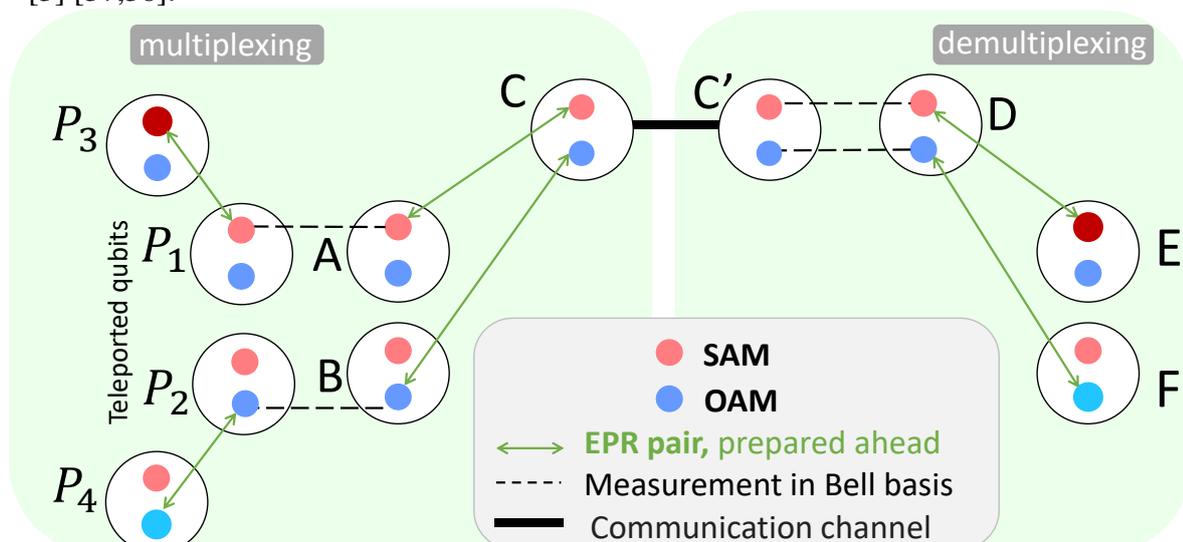

**Figure 3** Scheme for generating entangled photon pairs at a double rate. We start with initial states of the SAM of Photon $P_1$ entangled with the SAM of Photon $P_3$, and the OAM of Photon $P_2$ entangled with the OAM of Photon $P_4$. The green line represents entanglement between two DoFs, the black dashed line represents a measurement in Bell basis and the solid black line represents the communication channel. At the output we obtain the SAM of Photon E entangled with the SAM of Photon $P_3$ and the OAM of Photon F entangled with OAM of Photon $P_4$.

Thus far, we presented a technique that increased the transmission rate. However, since the quantum capacity also determines the rate of generating entangled pairs in remote locations, this protocol can be also valuable for entanglement generation for known quantum communication protocols such as quantum key distribution and super dense coding [39]. In the latter, we could use a Bell pair to transmit two classical bits over a single channel (a single qubit). Thus, this setting doubles the transmission rate. In our setting, we can multiply the transmission rate by a factor of 4 using the doubled generation of photon pairs. Figure 3 describes a similar setting for another purpose: generation of non-local entanglement between a transmitter and a receiver who are distant from each other, and here too our method facilitates entanglement generation at twice the rate. Namely, by sending only a single photon, we generate two non-local pairs of entangled photons at the destination. Our transmitter prepares ahead two entangled pairs: The SAM of Photons $P_3$ and $P_1$, and the OAM of Photons $P_2$ and $P_4$. Then, our transmitter locally teleports the SAM of photon $P_1$ and the OAM of photon $P_2$ to photon $C$ (as presented in Fig. 1(c)), and thus creating entanglement between the SAM of Photon $C$ with the SAM of Photon $P_3$, and the OAM of Photon $C$ with the OAM of Photon $P_4$. After the demultiplexing part, we end up with two separate pairs of entangled photons – Photon $E$ is entangled with Photon $P_3$ (in SAM DoF), and Photon $F$ is entangled with Photon $P_4$ (in OAM DoF). This feature of generating entangled pairs of photons at twice the rate can be beneficial and important in entanglement-assisted communication, increasing the channel capacity. Specifically, the entanglement-assisted classical capacity, which is the highest rate at which classical information can be transmitted from a sender to a receiver when they share an unlimited amount of noiseless entanglement. Thus, our protocol enables not only increasing the transmission rate of a quantum communication channel, but also increasing the rate of generating entangled pairs of qubits in remote locations.

To conclude, we proposed a hyperentanglement protocol for teleporting multiple entangled DoFs, multiplex them on a single photon, transmit the single photon at higher channel capacity, and eventually demultiplexing the information onto different photons which can be used for parallel processing of the quantum information. This system preserves the correlations between the different DoFs, despite being all carried by a single photon.

# Supplementary Information for

# Increasing Communication Rates Using Hyperentangled Photonic State

**Liat Nemirovsky[1], Uzi Pereg[2], and Mordechai Segev[1,2]***.

[1]*Solid State Institute, and Department of Physics, Technion-Israel Institute of Technology, Haifa 32000, Israel*
[2]*Department of Electrical Engineering, Technion, Haifa 3200003, Israel*
*\*msegev@technion.ac.il*

**This PDF file includes:**

Supplementary Text
Fig. S1
Fig. S2
Fig. S3
Fig. S4
References

## Section A: Calculating the Quantum Capacity

As explained in the main text, the quantum capacity is the ratio of information qubits per physical photon transmission. Every quantum channel $\mathcal{N}_{A \to B}$ has a Stinespring representation:

$$\mathcal{N}_{A \to B}(\rho) = \mathrm{Tr}_E(V \rho V^\dagger)$$

for every input density operator ρ on the Hilbert space $\mathcal{H}_A$, where the operator $V: \mathcal{H}_A \to \mathcal{H}_B \otimes \mathcal{H}_E$ satisfies $V^\dagger V = \mathbb{I}$. We refer to the systems $A$, $B$, and $E$ as belonging to Alice, Bob, and the environment, respectively.

Given a joint state $|\phi_{AA_1}\rangle$, where $A_1$ is an ancilla, denote the corresponding joint state of Bob, the environment, and the ancilla by

$$|\omega_{BEA_1}\rangle \equiv (V \otimes \mathbb{I})|\phi_{AA_1}\rangle$$

Then, the coherent information from Alice to Bob is defined as

$$I_C(A_1 \rangle B)_\omega \equiv H(\omega_B) - H(\omega_{BA_1}) = H(\omega_B) - H(\omega_E)$$

where $\omega_B$, $\omega_{BA_1}$, and $\omega_E$ are the reduced density operators of the respective systems, as $H(\omega) = -\mathrm{Tr}(\omega \, log(\omega))$ denotes the von Neumann entropy with respect to the density operator ω. The coherent information of the channel is defined as

$$I_c(\mathcal{N}) \equiv \max_{|\phi_{AA_1}\rangle} I_C(A_1 \rangle B)_\omega$$

where we optimize over all possible input states $|\phi_{AA_1}\rangle$.

Based on the Lloyd-Shor-Devetak result, a.k.a the LSD Theorem, the quantum capacity

$C_Q(\mathcal{N})$ of a given channel $\mathcal{N}_{A \to B}$ is given by the following formula,

$$C_Q(\mathcal{N}) \equiv \lim_{n \to \infty} \frac{1}{n} I_c(\mathcal{N}^{\otimes n})$$

In principle, to compute the quantum capacity for a general quantum channel, we need to compute the coherent information for the $n$-fold product channel $\mathcal{N}^{\otimes n}$, normalize by $n$, and then take $n$ to infinity. This could be difficult to compute. However, for the class of degradable channels, the characterization reduces to a much simpler formula:

$$C_Q(\mathcal{N}) \equiv I_c(\mathcal{N})$$

Intuitively, a quantum channel is degradable when the channel from Alice to Eve (the environment) is noisier than the channel from Alice to Bob.

In the standard case of an erasure channel, the quantum channel capacity is given by $C_Q = 1 - 2\epsilon$, for $0 \leq \epsilon \leq \frac{1}{2}$, where $\epsilon$ is the erasure rate. Otherwise, if $\frac{1}{2} \leq \epsilon \leq 1$, the quantum capacity is zero. Now, we calculate the quantum channel capacity of our case, where we transmit two qubits over a single transmission (a single photon contained two qubits).

To calculate the quantum channel capacity of our protocol, we calculate the mutual coherent information of our model, from $\boldsymbol{A_1 A_2}$ to $\boldsymbol{B_1 B_2}$, according to the scheme presented in Fig. S1:

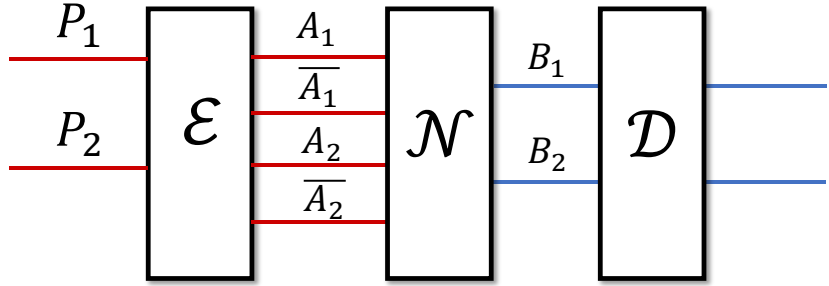

**Fig. S1**

In our calculation, $Z$ is an indicator: gives 0 if there wasn't erasing, and 1 if an erasure has occurred. The error (erasure) is orthogonal to the 2 transmitted qubits. $\mathcal{H}_B = \mathbb{C}^{\{0,1,e\}}$.

$$\begin{aligned} I_C(A_1 A_2 \rangle B_1 B_2)_\rho &= H(B_1 B_2)_\rho - H(A_1 A_2 B_1 B_2)_\rho = H(B_1 B_2 Z)_\rho - H(A_1 A_2 B_1 B_2 Z)_\rho \\ &= H(Z) + H(B_1 B_2 | Z) - [H(Z) + H(A_1 A_2 B_1 B_2 | Z)] \\ &= H(B_1 B_2 | Z) - H(A_1 A_2 B_1 B_2 | Z) = (*) \end{aligned}$$

We describe a Classical-quantum state as follows: $\rho_{ZB} = \sum_z p_z(z)|z\rangle\langle z| \otimes \rho_B^z$. Therefore:
$H(B|Z)_\rho = \sum_z p_z(z) H(\rho_B^z)$

$$(*) = p(z=0)H(\rho_{B_1B_2}^{z=0}) + p(z=1)H(\rho_{B_1B_2}^{z=1})$$
$$- [p(z=0)H(\rho_{A_1A_2B_1B_2}^{z=0}) + p(z=1)H(\rho_{A_1A_2B_1B_2}^{z=1})]$$
$$= (1-p)H\left(\frac{\mathbb{I}_{B_1}}{2} \otimes \frac{\mathbb{I}_{B_2}}{2}\right) + p \cdot H(|e\rangle\langle e| \otimes |e\rangle\langle e|)$$
$$- (1-p)H(\phi_{A_1\bar{A}_1} \otimes \phi_{A_2\bar{A}_2}) - p \cdot H\left(\frac{\mathbb{I}_{B_1}}{2} \otimes \frac{\mathbb{I}_{B_2}}{2} \otimes |e\rangle\langle e| \otimes |e\rangle\langle e|\right)$$

Now, for product states, $\rho_{AB} = \rho_A \otimes \rho_B$, we have $H(AB)_\rho = H(A)_\rho + H(B)_\rho$. Thus,

$$= (1-p) \cdot 2 \overbrace{H\left(\frac{\mathbb{I}}{2}\right)}^{=1} + p \cdot 2 \overbrace{H(|e\rangle\langle e|)}^{=0} - (1-p) \cdot 2 \overbrace{H(\phi)}^{=0} - p \cdot \left[ 2 \overbrace{H\left(\frac{\mathbb{I}}{2}\right)}^{=1} + 2 \overbrace{H(|e\rangle\langle e|)}^{=0} \right]$$

$$= 2(1-p) - 2p = \mathbf{2 - 4p}$$

As a final step, we show that the quantum capacity is increased while multiplexing more qubits on a single photon. One may notice that in the general case, where we multiplex $n$ degrees of freedom (DoFs), the quantum capacity will increase by a factor of n, following the same derivation:

$$I_C(A_1A_2 \dots A_n \rangle D_1D_2 \dots D_n)_\rho = H(D_1D_2 \dots D_n)_\rho - H(A_1A_2 \dots A_nD_1D_2 \dots D_n)_\rho$$
$$= H(D_1D_2 \dots D_n, Z)_\rho - H(A_1A_2 \dots A_n, D_1D_2 \dots D_n, Z)_\rho$$
$$= H(Z) + H(D_1D_2 \dots D_n|Z) - [H(Z) + H(A_1A_2 \dots A_n, D_1D_2 \dots D_n|Z)]$$
$$= H(D_1D_2 \dots D_n|Z) - H(A_1A_2 \dots A_n, D_1D_2 \dots D_n|Z) =$$

$$= p(z=0)H(\rho_{D_1D_2 \dots D_n}^{z=0}) + p(z=1)H(\rho_{D_1D_2 \dots D_n}^{z=1})$$
$$- [p(z=0)H(\rho_{A_{SAM}B_{OAM},D_1D_2 \dots D_n}^{z=0})$$
$$+ p(z=1)H(\rho_{A_{SAM}B_{OAM},D_1D_2 \dots D_n}^{z=1})]$$

$$= (1-p)H\left(\frac{\mathbb{I}_{D_1}}{2} \otimes \frac{\mathbb{I}_{D_2}}{2} \dots \otimes \frac{\mathbb{I}_{D_n}}{2}\right) + p \cdot H(|e\rangle\langle e| \otimes |e\rangle\langle e| \dots \otimes |e\rangle\langle e|)$$
$$- (1-p)H(\phi_{A_1\bar{A}_1} \otimes \phi_{A_2\bar{A}_2} \dots \otimes \phi_{A_n\bar{A}_n}) - p$$
$$\cdot H\left(\frac{\mathbb{I}_{D_1}}{2} \otimes \frac{\mathbb{I}_{D_2}}{2} \otimes \dots \frac{\mathbb{I}_{D_n}}{2} \otimes |e\rangle\langle e| \otimes |e\rangle\langle e| \dots \otimes |e\rangle\langle e|\right)$$

$$= (1-p) \cdot n \overbrace{H\left(\frac{\mathbb{I}}{2}\right)}^{=1} + p \cdot n \overbrace{H(|e\rangle\langle e|)}^{=0} - (1-p) \cdot n \overbrace{H(\phi)}^{=0} - p$$

$$\cdot \left[ n \overbrace{H\left(\frac{\mathbb{I}}{2}\right)}^{=1} + n \overbrace{H(|e\rangle\langle e|)}^{=0} \right] = n(1-p) - np = (1-2p)n$$

This means, that our technique with multiplexing qubits on a single photon - instead of transmitting several photons while each is encoded with a single qubit - has a higher quantum channel capacity. In fact, we can increase the quantum channel capacity more than twice, comparing to the standard case, if we multiplex more than two DoF.

## Superactivation: The Quantum Capacity of a Product Channel May Surpass the Sum-Rate

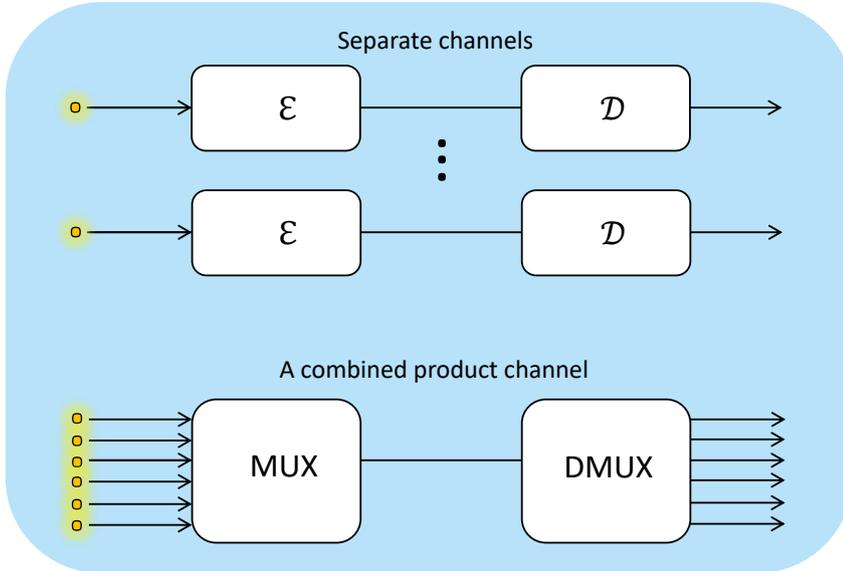

**Fig. S2 Coding for the Combined Product Channel vs. Coding for Each Channel Separately**

As discussed in the main text, the quantum capacity of the combined (product) channel can be larger than the sum of the quantum capacities of each individual channel, .i.e., encoding several qubits on a single photon (a single transmission) can in principle enlarge the quantum capacity, comparing to a case of encoding a single qubit on a photon.

In addition, as mentioned in the main text, one can even achieve super activation- where the quantum channel capacity of each channel (for each DoF) is zero, but the quantum channel capacity of the combined product channel takes a positive value [1]. In particular, this is the case when combining a symmetric erasure channel (that is, an error rate of $\epsilon = \frac{1}{2}$), and an entanglement binding channel [2]. This feature is unique to quantum communication, and does not occur in classical communication.

### Section B: Simulating states with erasure channel

The simulation we used in Fig. 2 in the main text is performing the following quantum circuit:

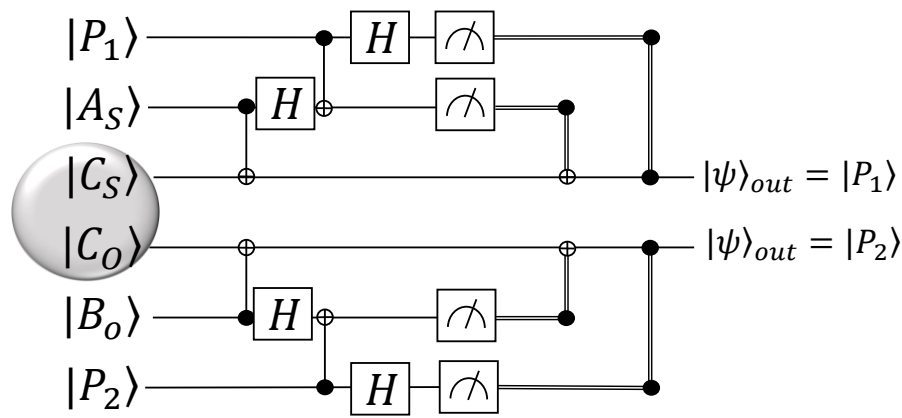

**Fig. S3 Quantum Circuit for the proposed protocol**

We simulate the circuit presented in Fig. S3 and reconstructed different initial states, generated randomly. After each operation we performed an erasure error, where we check the reconstruction of the states for different error rates (as presented in Fig. 2(b) in the main text).

### Section C: Increasing the Hilbert space for higher communication rates

We distinguish between two strategies to increase the Hilbert space: the first is adding $k$ DoFs as 2-level systems (i.e, adding more qubits); and the second is adding just one DoF (total of 2 DoFs) but DoFs are an n-level systems, where $n = 2^k$ (i.e., adding a qudit). From a purely mathematical perspective, the two options are equivalent. Both actions enlarge the Hilbert space, and enable multiplexing a greater amount of information on a single photon. Figure R1 represents the two cases.

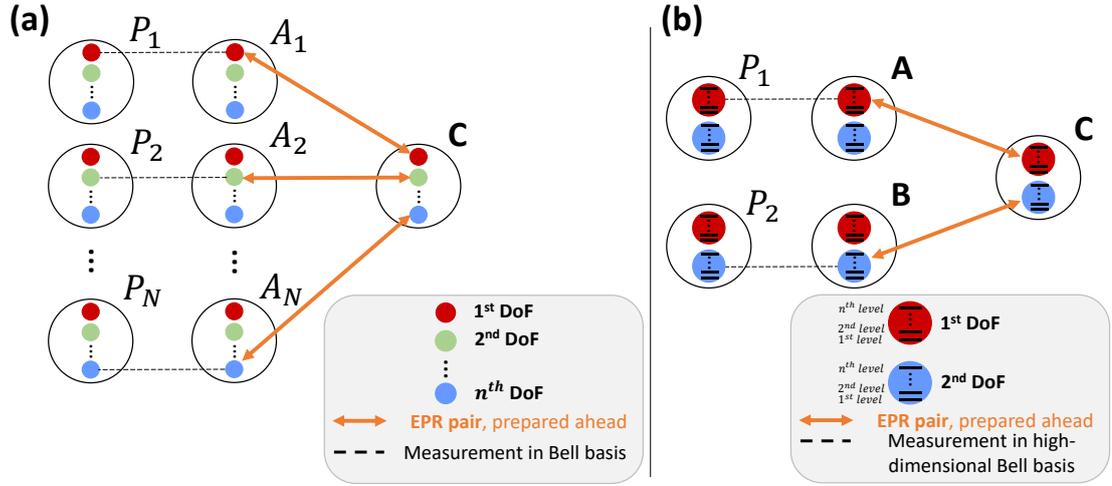

**Fig. S4 Two cases for increasing the Hilbert space in our protocol. (a) Adding n DoFs, where each DoF is a two-level system. (b) Perform the protocol with just 2 DoFs, but now each DoF is n-level system.**

In the first scheme (Fig. S4(a)), the complete Bell state analysis would still require to detect a single outcome (out of four possible measurement outcomes). That is, we would still need to perform a Bell state measurement on two qubits, but now perform that Bell state measurement N times (See Fig. S4(a)). The difficulty in this case lies in the initial state we need to prepare in order to perform multiplexing. This would require a hyperentangled state with different DoFs (2-level systems) encoded on photons $A_1, \ldots A_N$ and photon C.

In the second scheme (Fig. S4(b)), the issue of a complete Bell state analysis becomes more complicated. It requires now a d-dimensional Bell state analysis, as in [3-4]. For example, consider a bipartite state with a DoF which is a 3-level system. An example for a bipartite entangled state of this kind would be: $|\psi\rangle = \frac{1}{\sqrt{3}}(|00\rangle + |11\rangle + |22\rangle)$, while there are other 8 possible states, to form a complete orthonormal basis of the 3d bipartite Hilbert space. In the general case, for an n-level bipartite system, there exists $n^2$ high-dimensional Bell states. This means that, to be able to perform a complete Bell state analysis, we would have to be able to detect one (out of $n^2$ possible measurement outcomes). It was shown [10] that in theory, it is impossible to discriminate two-photon high-dimensional Bell states with linear optics for dimensions n ≥ 3. However, Y. H Luo et al [3] showed that such a high-dimensional Bell state analysis is achievable, even with $n \geq 3$, using a multiport beam splitter with n-input–n-output all-to-all connected ports.

Altogether, in the first scheme the dimension of the Hilbert space increases exponentially, and the Bell state analysis is more straightforward, compared to the second scheme where a high-dimensional Bell state analysis is required. Moreover, in the second case the space dimension increases linearly (not exponentially). This makes the first option favorable when trying to increase the communication rate in our protocol.

The concept of using hyperentanglement for increasing the channel capacity of a quantum channel is not limited in the number of DoFs. However, technology always presents practical considerations that limit the number of DoFs in both methods. Specifically, using the first approach we will have a limit on the initial hyperentangled state we can prepare, and, using the second approach - the dimension of the qudits we can add to our protocol is limited because our protocol requires an efficient way to detect a single Bell state. Nonetheless, by adding more DoFs as qubits (i.e., two level systems), we would still be able to perform at least the multiplexing part (although the initial state would be a larger hyperentangled state, which is generally harder to prepare).